
\input phyzzx
\input tables \date={1993}
\rightline {March 1993} \rightline {SUSX--TH--93/11.}
\rightline {(Revised)}
\title {Yukawa Couplings involving Excited Twisted Sector States
for ${\bf Z}_M\times {\bf Z}_N$ Orbifolds}
\author{D. Bailin${a}$, \ A. Love${b}$ \ and \ W.A.
Sabra${b}$}
\address {${a}$School of Mathematical and Physical
Sciences,\break
University of Sussex, \break Brighton U.K.}
\address {${b}$Department of Physics,\break
Royal Holloway and Bedford New College,\break
University of London,\break
Egham, Surrey, U.K.}

\abstract{A study is made for ${\bf Z}_M\times {\bf Z}_N$ orbifolds
of the modification of the form of the twisted sector
Yukawa couplings when some of the states involved are excited
twisted sectors rather than twisted sector ground states.}
\endpage
\REF\one{L. Dixon, J. A. Harvey, C. Vafa and E. Witten, Nucl. Phys.
B261 (1985) 678; B274 (1986) 285.}
\REF\two{ A. Font, L. E. Ibanez,
F. Quevedo and A. Sierra, Nucl. Phys. B331 (1991) 421.}
\REF\three{L. Dixon, D. Friedan, E. Martinec and S. Shenker, Nucl.
Phys. B282 (1987) 13.}
\REF\four{S. Hamidi, and C. Vafa, Nucl. Phys. B279 (1987) 465.}
\REF\five{ L. E. Ibanez, Phys. Lett. B181 (1986) 269.}
\REF\six{J. A. Casas and C. Munoz, Nucl. Phys. B332 (1990) 189}
\REF\seven {J. A. Casas, F. Gomez and C. Munoz,  Phys. Lett.
B251(1990) 99}
\REF\eight {J. A. Casas, F. Gomez and C. Munoz,  CERN preprint,
TH6194/91.}
\REF\nine{T. Kobayashi and N. Ohtsubu, Kanazawa
preprint, DPKU--9103.}
\REF\ten {T. T. Burwick, R. K. Kaiser and H. F.
Muller, Nucl. Phys. B355 (1991) 689}
\REF\eleven{J. Erler, D. Jungnickel and J. Lauer, Phys. Rev D45
(1992) 3651; S. Stieberger, D. Jungnickel, J. Lauer
and M. Spalinski, Mod. Phys. Lett. A7 (1992) 3059; J. Erler,  D.
Jungnickel, M. Spalinski and S. Stieberger, preprint,
MPI--Ph/92--56.}
\REF\twelve{D. Bailin, A. Love and W. A. Sabra,
Mod. Phys. Lett A6 (1992) 3607; D. Bailin, A. Love and W. A. Sabra,
Sussex preprint, SUSX-TH-92/17, to be published in Nucl. Phys. B.}
\REF\thirteen{S. Stieberger, preprint, TUM--TH--151/92.}
\REF\fourteen{V. S. Kaplunovsky, Nucl. Phys. B307 (1988) 145; L. J.
Dixon, V. S. Kaplunovsky and J. Louis,  Nucl. Phys. B355 (1991) 649;
J. P. Derenddinger, S. Ferrara, C. Kounas and F. Zwirner, Nucl.
Phys. B372 (1992) 145, Phys. Lett. B271 (1991) 307.}
\REF\fifteen{I. Antoniadis, J. Ellis, R. Lacaze and D. V.
Nanopoulos, Phys. Lett. B268 (1991) 188; S. Kalara, J. L. Lopez
and D. V. Nanopoulos,  Phys. Lett. B269 (1991) 84.}
\REF\sixteen {L. E. Ibanez and D. L\"ust
and G. G. Ross, Phys. Lett. B272 (1991) 251; D. Bailin and A. Love,
Phys. Lett. B278 (1992) 125.}
\REF\seventeen{L. E. Ibanez and D.
L\"ust, Nucl. Phys. B382 (1992) 305.}
\REF\eighteen{D. Bailin and A. Love,
Phys. Lett. B292 (1992) 315.}
\REF\nineteen{A. Font, L. E.
Ibanez, H-P. Nilles  and F. Quevedo,  Nucl. Phys. B307 (1988)
109.}
\REF\twenty{A.A. Belavin, A.M. Polyakov and A.B. Zamolodchikov,
Nucl. Phys. B241 (1984) 333.}
A knowledge of the Yukawa couplings for orbifold compactified
string theory models [\one, \two] will be required for a
comparison of such models with observation. In particular,
the exponential dependence of Yukawa
couplings on moduli which can occur when all the states
involved are in twisted sectors [\three, \four]
may have a bearing on
hierarchies of quark and lepton masses [\five].

Twisted sector Yukawa couplings have already been investigated for
both the ${\bf Z}_N$ orbifolds [\three--\eleven] and  for the
${\bf Z}_M\times {\bf Z}_N$ orbifolds [\twelve--\thirteen].
However, the discussion has been mostly limited to couplings
involving only twisted sector ground states, though an outline has
been given of how twisted sector excited states might be included
[\three, \four].
Here we shall extend the discussion to Yukawa couplings involving
twisted sector excited states of  ${\bf Z}_M\times {\bf Z}_N$
Coxeter orbifolds. It is particularly important to be able to
include these excited states in view of the fact that string loop
threshold corrections to gauge coupling constants
[\fourteen,\fifteen] consistent with their low energy values have
so far always involved modular weights for quarks and leptons
requiring the use of excited twisted sector states.

The ${\bf Z}_M\times {\bf Z}_N$ orbifolds under consideration are
those for which the point group is realised in the simplest
possible way in terms of the Coxeter elements of Lie algebra
root lattices as discussed in the second reference of [\twelve] and
summarized in table 1.  We are interested in massless states in the twisted
sect
 ors of
these orbifolds with the $SU(3)\times SU(2)\times U(1)$ quantum
numbers of quarks, leptons and electroweak Higgses. Possible
(fractional) oscillator numbers $\tilde N$ for massless states with
these gauge group quantum numbers are bounded by [\seventeen]
$${\tilde N}\le a_L-{3\over5},\qquad {\hbox{for}}\ Q, \ u_c\
{\hbox {and}}\ e_c\eqn\mi$$ and by
$${\tilde N}\le a_L-{2\over5},\qquad {\hbox{for}}\ L,\ d_c\
{\hbox{and}}\ H,\eqn\mig$$ where $a_L$ is the left mover normal
ordering constant for the twisted sector in question. The inequality
in \mi\ and \mig\ reflects the generic occurence in orbifold models,
before spontaneous symmetry breaking, of extra $U(1)$ gauge fields to
which the quarks and leptons couple. The bounds \mi\ and \mig\
put tight constraints on the left movers bosonic oscillators that
can be deployed in the construction of massless twisted sector
states.

Allowed Yukawa couplings involving excited twisted sector states
are restricted by the observation [\four, \nineteen]
that the discrete symmetries of the 2-dimension-\hfill\break al
sub-lattices of the 6-dimensional compact manifold are left unbroken
by the construction of the orbifold. If $Xi,$ $i=1,2,3,$ are the
complex coordinates defining the 6-dimensional compact manifold and a
discrete symmetry acting in the $i$-th complex plane is of order
$P$, then correlation functions involving $\big(\partial_{\bar
z}Xi\big)m \big(\partial_{\bar w}{\bar X}i\big)n$ are allowed
only if  $$m-n=0\ \ \ ({\hbox {mod}}\ P).\eqn\migr$$

Yukawa couplings involving only twisted sector ground states are
determined by bosonic twist field correlation functions
[\three, \four] of the type
$${\bf Z}=\prod_{i=1}3{\cal Z}_i,\eqn\migra$$
with
$${\cal Z}_i=\langle\sigma_\alphai(z_1, \bar z_1)
\sigma_{\beta}i(z_2, \bar z_2)\sigma_{\gamma}i(z_3,
\bar z_3)\rangle,\eqn\migran$$
where the subscripts $\alpha$, $\beta$ and $\gamma$ on the twist
fields are the point group elements for the three twisted sectors
involved, and because of the point group selection rule
$$\gamma=\big(\alpha\beta\big){-1}.\eqn\migrane$$
If excited twisted sector states are involved then excited twist
fields ${\tilde\tau}$ and ${\tilde\tau}'$ are required which are defined by the
operator product expansions [\three,\four]
$$\partial_{\bar z}X \sigma_\alpha(w, \bar w)\sim (\bar z-\bar
w){-{\eta}_\alpha}{{\tilde\tau}_\alpha}'(w,\bar w)+....\eqn\me$$
and
$$\partial_{\bar z}{\bar X} \sigma_\alpha(w, \bar w)\sim (\bar
z-\bar w){-(1-{\eta}_\alpha)}{{\tilde\tau}_\alpha}(w,\bar w)+....\eqn\med$$
where in the $\alpha$ twisted sector the coordinates $Xi$ are
twisted by $e{-2\pi i{\eta_\alphai}}$ and the index $i$
labelling the $i$-th complex plane of the 6-dimensional compact
manifold has been suppressed in \me\ and \med.
(The excited twist fields ${\tilde\tau}_\alpha$ and ${{\tilde\tau}_\alpha}'$
are
associated with the excited twisted sector states created by a
single bosonic oscillator corresponding to moding $\eta_\alpha$ or
$1-\eta_\alpha$. This is the only situation that turns out to be
relevant here.)
Then, we are interested in correlation functions \migra\ where at
least one of the factors ${\cal Z}_i$ involves excited twist fields
and is of the form
$${\Big({\cal Z}_i3\Big)}_{excited}=
\langle{\tilde\tau}_\alphai(z_1, \bar z_1)
{\tilde\tau}'{}i_\beta(z_2, \bar z_2)\sigma_{\gamma}i(z_3,
\bar z_3)\rangle,\eqn\medi$$
up to  permutations of $\alpha$, $\beta$ and $\gamma$. If the point
group element $\gamma$ leaves the $i$-th complex plane unrotated
then $\sigma_{\gamma}i$ is trivial and \medi\ reduces to the
excited two-point function
$${\Big({\cal Z}_i2\Big)}_{excited}=\langle{{\tilde\tau}_\alpha}i(z_1,
\bar z_1) {\tilde\tau}'{}i_{\alpha{-1}} (z_2, \bar z_2)\rangle,\eqn\medic$$
which can be normalised to 1 apart from factors of $(z_1-z_2)$
and $(\bar z_1-\bar z_2)$ determined by $SL(2,C)$ invariance. (The
same remark applies to two-point functions where the excited states
are created by a product of two bosonic oscillators.) Thus,
non-trivial modifications of the Yukawa couplings due to the use of
excited twist fields only occur when $\sigma_{\gamma}i$  is
non-trivial in \medi. As an illustration, the sectors of the   ${\bf
Z}_3\times {\bf Z}_6$ orbifold for which this happens are displayed
in table 2, using the notation $T_{pq}$ to denote the
$\thetap\omegaq$ twisted sector.

Normalization of the excited twist fields is necessary in order to
create normalized states. The twisted sector mode expansion for $X$
and $\bar X$ suggests that the normalization factors for
${\tilde\tau}_\alpha$ and ${\tilde\tau}'_\alpha$ should be
essentially $\big(2\eta_\alpha\big){-{1\over2}}$ and
$\big(2(1-\eta_\alpha)\big){-{1\over2}}$, respectively. This can be
checked in detail by considering the excited two-point function
\medic\ for $\beta=\alpha{-1}$ which can be derived from the
correlation function $${\bar g}(\bar z,\bar w)={-{1\over2}}
{\langle\partial_{\bar z}{X}\partial_{\bar w}{\bar
X}\sigma_\alpha(z_1, \bar z_1)\sigma_{\alpha{-1}}(z_2,
\bar z_2\rangle\over\langle\sigma_\alpha(z_1, \bar
z_1)\sigma_{\alpha{-1}}(z_2,
\bar z_2\rangle},\eqn\medici$$
where we have suppressed the index $i$. With the aid of the operator
product expansions \me\ and \med, the excited twist field two-point
function is related to ${\bar g}(\bar z,\bar w)$ by
$$\eqalign{\langle{\tilde\tau}_\alpha(z_1, \bar
z_1){{\tilde\tau}'}_{\alpha{-1}
 }(z_2,
\bar z_2)\rangle=&-2\langle\sigma_\alpha (z_1, \bar
z_1)\sigma_{\alpha{-1}}(z_2, \bar z_2)\rangle\cr
&\lim_{w\rightarrow z_1\atop z\rightarrow z_2}(\bar z-\bar
z_2){1-\eta_\alpha}(\bar w-\bar z_1){1-\eta_\alpha}
{\bar g}(\bar z,\bar w).}\eqn\medicin$$
The correlation function ${\bar g}(\bar z,\bar w)$ has has the form
[\three]
$$\eqalign{&{\bar g}(\bar z,\bar w)=({\bar z}-{\bar
z}_1){-\eta_\alpha} ({\bar z}-{\bar z}_2){-(1-\eta_\alpha)}
({\bar w}-{\bar z}_1){-(1-\eta_\alpha)}
({\bar w}-{\bar z}_2){-\eta_\alpha}\cr
&\Big({(1-\eta_\alpha)(({\bar
w}-{\bar z}_1)({\bar z}-{\bar z}_2)+\eta_\alpha({\bar z}-{\bar
z}_1)({\bar w}-{\bar z}_2)\over (\bar z-\bar
w)2}\Big).}\eqn\medicine$$ Combining \medicin\ and \medicine\ gives
$$\langle{\tilde\tau}_\alpha(z_1){{\tilde\tau}'}_{\alpha{-1}}(z_2)\rangle=
2\eta_\alpha(-1){-\eta_\alpha} ({\bar z}_2-{\bar
z}_1){-2\eta_\alpha}{\mid
z_2-z_1\mid}{-4h_\alpha}.\eqn\ma$$
where $h_\alpha = {1\over2}\eta_\alpha(1-\eta_\alpha)$ is the conformal
weight of $\sigma_\alpha$.  The factors of
$({\bar z}_2-{\bar z}_1)$ and $(z_2-z_1)$ are to be expected
because of $SL(2,C)$ invariance (see for example, ref [20]).
This confirms the remark at the beginning of the paragraph that the
normalization factors for ${\tilde\tau}_\alpha$ and
${{\tilde\tau}'}_\alpha$ are essentially
$\Big(2\eta_\alpha\Big){-{1\over2}}$ and
$\Big(2(1-\eta_\alpha)\Big){-{1\over2}}$, respectively, but also
makes it clear that there is a factor of $(-1){-\eta_\alpha}$ due
to the fact that the excited twist fields have a non-zero spin. Such
a factor is absorbed in the definition of the $in$ and $out$ states.

The
three-point function \medi\ can be evaluated directly apart from an
overall moduli independent normalization factor and this overall
normalization can be determined by factorizing a four-point
function. Using the operator product expansions \me\ and \med,
$$\eqalign{{\Big({\cal Z}3\Big)}_{excited}=&\lim_{w\rightarrow
z_1\atop z\rightarrow z_2} (\bar z-{\bar z}_2){\eta_\beta} (\bar
w-{\bar z}_1){1-\eta_\alpha} \cr &\langle\partial_{\bar
z}{X}\partial_{\bar w}{\bar X} {\sigma_\alpha}(z_1, \bar
z_1){\sigma_{\beta}}(z_2, \bar z_2)\sigma_{(\alpha\beta){-1}} (z_3,
\bar z_3)\rangle.}\eqn\maxwel$$
where the index $i$ labelling the $i$th complex plane has been
suppressed in this and subsequent equations.  Separating $X$ into a classical
pa
 rt and a
quantum part $$X=X_{cl}+X_{qu},\eqn\maxwell$$ we have
$$\eqalign{&\langle\partial_{\bar z}{X}\partial_{\bar w}{\bar
X}\sigma_\alpha(z_1, \bar z_1)\sigma_{\beta}(z_2, \bar
z_2)\sigma_{(\alpha\beta){-1}}(z_3, \bar z_3)\rangle=\cr
&\sum_{X_{cl}}e{-S_{cl}} {\langle\partial_{\bar z}{X}_{qu}\partial_{\bar
w}{\bar X}_{qu}\rangle}_{{3-twists}} +
\sum_{X_{cl}}e{-S_{cl}}
{\partial_{\bar z}{X}}_{cl}\partial_{\bar w}{\bar
X}_{cl}{\cal Z}_{qu}3,}\eqn\h$$
where
$$\eqalign{{\langle\partial_{\bar z}{X}_{qu}\partial_{\bar w}{\bar
X}_{qu}\rangle}_{{3-twists}}=&\int {\cal D}X_{qu}e{-S_{qu}}
\partial_{\bar z}{X}_{qu}\partial_{\bar w}{\bar
X}_{qu}\cr &\sigma_\alpha(z_1, \bar z_1)\sigma_{\beta}(z_2,
\bar z_2)\sigma_{(\alpha\beta){-1}}(z_3, \bar
z_3),}\eqn\he$$
$${\cal Z}_{qu}3=
\int {\cal D}X_{qu}e{-S_{qu}}
\sigma_\alpha(z_1, \bar z_1)\sigma_{\beta}(z_2,
\bar z_2)\sigma_{(\alpha\beta){-1}}(z_3, \bar
z_3)\eqn\hel$$
and
$$S={1\over\pi}\int d2z
\Big(\partial_{z}{X}\partial_{\bar z}{\bar
X}+\partial_{\bar z}{X}\partial_{z}{\bar
X}\Big).\eqn\hell$$
Continuing to suppress the index $i$ on $Xi$, the classical fields
consistent with the operator product expansions \me\ and \med\ have
derivatives of the form
$$\partial_{\bar z}X_{cl}=d (\bar z-{\bar z}_1){-\eta_\alpha}
(\bar z-{\bar z}_2){-\eta_\beta}
(\bar z-{\bar z}_3){-(1-\eta_\alpha-\eta_\beta)},\eqn\p$$
and$$\partial_{\bar w}{\bar X}_{cl}=a (\bar
w-{\bar z}_1){-(1-\eta_\alpha)} (\bar w-{\bar
z}_2){-(1-\eta_\beta)} (\bar
w-{\bar z}_3){-(\eta_\alpha+\eta_\beta)}.\eqn\pl$$
The constant $d$ must be chosen to be zero for an acceptable
classical solution because the classical action is otherwise
divergent. Consequently, the second term in \h\ vanishes, and the
moduli dependence of ${\Big({\cal Z}3\Big)}_{excited}$, which is
contained in  ${\sum}_{X_{cl}}e{-S_{cl}}$, is exactly the same as
for the three-point function with unexcited twist fields.

Determination of the overall normalization of the three-point
function, which depends on the twisted sectors involved, can be
achieved by considering the four-point function,
$${\Big({\cal
Z}4\Big)}_{excited}=\langle{\tilde\tau}_{\alpha{-1}}(z_1,\bar z_1 )
{\sigma_\alpha}(z_2,\bar z_2 ){{\tilde\tau}'}_{\beta{-1}}(z_3, \bar z_3 )
\sigma_{\beta}(z_4, \bar z_4 )\rangle.\eqn\pla$$  With the aid of the
operator product expansions \me\ and \med\ this can be written as
$$\eqalign{{\Big({\cal Z}4\Big)}_{excited}=&\lim_{w\rightarrow
z_1\atop z\rightarrow z_3} (\bar w-{\bar z}_1){\eta_\alpha}
(\bar z-{\bar z}_3){1-\eta_\beta}\cr
&\langle\partial_{\bar z}{X}\partial_{\bar w}{\bar
X}\sigma_{\alpha{-1}}(z_1,{\bar z}_1)
{\sigma_\alpha}(z_2, {\bar z}_2){\sigma}_{\beta{-1}}(z_3, {\bar
z}_3) \sigma_{\beta}(z_4, {\bar z}_4)\rangle.}\eqn\plan$$

Separating $X$ into a classical part and quantum part as in
\maxwell,
$$\eqalign{&\langle\partial_{\bar z}{X}\partial_{\bar w}{\bar
X}\sigma_{\alpha{-1}}(z_1,{\bar z}_1)
{\sigma_\alpha}(z_2, {\bar z}_2){\sigma}_{\beta{-1}}(z_3, {\bar
z}_3) \sigma_{\beta}(z_4, {\bar
z}_4)\rangle=\cr &\sum_{X_{cl}}e{-S_{cl}} {\langle\partial_{\bar
z}{X}_{qu}\partial_{\bar w}{\bar X}_{qu}\rangle}_{{4-twists}} +
\sum_{X_{cl}}e{-S_{cl}}
{\partial_{\bar z}{X}}_{cl}\partial_{\bar w}{\bar
X}_{cl}{\cal Z}_{qu}4,}\eqn\plant$$
where
$$\eqalign{&{\langle\partial_{\bar z}{X}_{qu}\partial_{\bar w}{\bar
X}_{qu}\rangle}_{{4-twists}}=\cr
& \int {\cal D}X_{qu}e{-S_{qu}}
\partial_{\bar z}{X}_{qu}\partial_{\bar w}{\bar
X}_{qu}
\sigma_{\alpha{-1}}(z_1,{\bar z}_1)
{\sigma_\alpha}(z_2, {\bar z}_2){\sigma}_{\beta{-1}}(z_3, {\bar
z}_3) \sigma_{\beta}(z_4, {\bar z}_4)\rangle}\eqn\plants$$
and
$${\cal Z}_{qu}4=
\int {\cal D}X_{qu}e{-S_{qu}}
\sigma_{\alpha{-1}}(z_1,{\bar z}_1)
{\sigma_\alpha}(z_2, {\bar z}_2){\sigma}_{\beta{-1}}(z_3, {\bar
z}_3) \sigma_{\beta}(z_4, {\bar z}_4).\eqn\le$$
The first term in \plant\ can be evaluated from
$${\bar h}(\bar z,\bar w)=
{-{1\over2}}{{\langle\partial_{\bar
z}{X}_{qu}\partial_{\bar w}{\bar
X}_{qu}\rangle}_{{4-twists}}\over{\cal Z}_{qu}4} .\eqn\leb$$  Using
the operator product expansion  $${-{1\over2}}\partial_{\bar
z}{X}\partial_{\bar w}{\bar X}\sim {1\over (\bar z-\bar w)2}+
 \hbox{finite}\eqn\leba$$ together with the operator product
expansions \me\ and \med, we require ${\bar h}(\bar z,\bar w)$ to
satisfy $${\bar h}(\bar z,\bar w)\sim{1\over (\bar z-\bar
w)2}+{\hbox{finite}},\qquad {\bar z}\rightarrow {\bar w}\eqn\leban$$
$${\bar h}(\bar z,\bar w)\sim (\bar z-{\bar z}_1){-(1-\eta_\alpha)},
\qquad {\bar z}\rightarrow {\bar z}_1 \eqn\lebano$$
and
$${\bar h}(\bar z,\bar w)\sim (\bar w-{\bar z}_1){-\eta_\alpha},
\qquad {\bar w}\rightarrow {\bar z}_1 \eqn\lebanon$$
with similar conditions for $\bar z\rightarrow {\bar z}_2,
{\bar z}_3,{\bar z}_4$ and ${\bar w}\rightarrow {\bar z}_2,{\bar
z}_3,{\bar z}_4$.
Following the methods of [\three, \ten], we arrive at
$$\eqalign{&\lim_{w\rightarrow z_1\atop z\rightarrow z_3}
(\bar w-{\bar z}_1){\eta_\alpha}
(\bar z-{\bar z}_3){1-\eta_\beta}
{\bar h}(\bar z,\bar w)=\cr &(-1){3\eta_\alpha-\eta_\beta}(\bar
x){\eta_\alpha}(1-\bar x){-{\eta_\alpha}}
\Big((\eta_\alpha-\eta_\beta)-(1-\bar x)\partial_{\bar x}log
 I (x,\bar x)\Big),}\eqn\en$$
where we used $SL(2,C)$ invariance to set
$$z_1=0,\ z_2=x,\ z_3=1,\ z_4=z_\infty,\eqn\eng$$
and
$$I (x,\bar x)=J_2\bar G_1(\bar x){H}_2(1-x)+
J_1{G}_2(x){\bar H}_1(1-\bar x),\eqn\engla$$
with
$$J_1={\Gamma({\eta_\alpha})\Gamma(1-{\eta_\beta})
\over\Gamma(1+{\eta_\alpha}-{\eta_\beta})},\qquad
J_2={\Gamma(1-{\eta_\alpha})\Gamma({\eta_\beta})
\over\Gamma(1+{\eta_\beta}-{\eta_\alpha})},\eqn\engla$$
$$G_1(x)=F({\eta_\alpha},1-{\eta_\beta};1;x),\qquad
G_2(x)=F(1-{\eta_\alpha},{\eta_\beta};1;x)\eqn\englan$$
and
$$H_1(x)=
F({\eta_\alpha},1-{\eta_\beta};1+{\eta_\alpha}-{\eta_\beta};x),\qquad
H_2(x)=
F(1-{\eta_\alpha},{\eta_\beta};1+{\eta_\beta}-{\eta_\alpha};x).
\eqn\england$$
To factorize ${\Big({\cal Z}4\Big)}_{excited}$ into a product of
Yukawa couplings we shall, as in the case with only unexcited twist
fields [\three,\ten] have to take the limit $z_2\rightarrow z_4,$
i.e., $x\rightarrow z_\infty$. In this limit, we find from the
asymptotic behaviour of the hypergeometric functions $F,$ that
$$\eqalign{&\lim_{{x\rightarrow z_\infty \atop w\rightarrow
z_1}\atop z\rightarrow z_3} (\bar w-{\bar z}_1){\eta_\alpha}
(\bar z-{\bar z}_3){1-\eta_\beta}
{\langle\partial_{\bar z}{X}_{qu}\partial_{\bar w}{\bar
X}_{qu}\rangle}_{{4-twists}}\cr
=&\lim_{{x\rightarrow z_\infty \atop w\rightarrow z_1}\atop
z\rightarrow z_3} -2{\cal Z}_{qu}4
(\bar w-{\bar z}_1){\eta_\alpha}
(\bar z-{\bar z}_3){1-\eta_\beta}
{\bar h}(\bar z,\bar w)\cr
=&\lim_{x\rightarrow z_\infty}
2\eta_{\beta}(-1){2\eta_\alpha-\eta_\beta}{\cal
Z}_{qu}4\qquad {\hbox{for}}\ \eta_\alpha< 1-\eta_\beta,\cr &
\lim_{x\rightarrow
z_\infty}2(1-\eta_{\alpha})(-1){2\eta_\alpha-\eta_\beta}{\cal Z}_{qu}4\qquad
{\hbox{for}}\ \eta_\alpha>1-\eta_\beta.}\eqn\fr$$ The second term in \plant\
can
be evaluated by writing [\ten], consistently with the operator product
expansions \me\ and \med, $$\partial_{\bar z}X_{cl}=c{(\bar
z)}{\eta_{\alpha{-1}}}{(\bar z-\bar x)}{-\eta_{\alpha}} {(\bar
z-1)}{\eta_{\beta{-1}}} {(\bar z-{\bar z}_\infty)}{-\eta_{\beta}}\eqn\fra$$
and
$$\partial_{\bar z}{\bar X}_{cl}={\bar b}{(\bar
z)}{-\eta_{\alpha}} {(\bar z-\bar x)}{\eta_{\alpha{-1}}}
{(\bar z-1)}{-\eta_{\beta}}{(\bar z-\bar
z_\infty)}{\eta_{\beta{-1}}}\eqn\fran$$ Then, $c$ and $b$ can be
derived in terms of hypergeometric functions from the independent
global monodromy conditions, as in the third reference of
[\eleven], and taking the limit $x\rightarrow \infty$ we find that
the second term in \plant\ vanishes.

Now, with the aid of \fr, we
find that for $x\rightarrow z_\infty$,
$$\eqalign{{\Big({\cal
Z}4\Big)}_{excited}=&\lim_{x\rightarrow
z_\infty}2\eta_{\beta}(-1){2\eta_\alpha-\eta_\beta}{\cal Z}_{qu}4
\sum_{X_{cl}}e{-{S_{cl}}},\qquad {\hbox{for}}\ \eta_\alpha<
1-\eta_\beta,\cr &
\lim_{x\rightarrow
z_\infty}2(1-\eta_{\alpha})(-1){2\eta_\alpha-\eta_\beta}{\cal Z}_{qu}4
\sum_{X_{cl}}e{-{S_{cl}}}\qquad {\hbox{for}}\
\eta_\alpha>1-\eta_\beta.}\eqn\franc$$ The factorization into a
product of three-point functions now proceeds much as in the absence
of excited twist fields [\ten,\eleven], and we conclude that
$$\eqalign{{\langle{\tilde\tau}_{\alpha{-1}}(0){{\tilde\tau}'}_{\beta{-1}}(1)
\sigma_{\alpha\beta}(z_\infty)\rangle\over
\langle\sigma_{\alpha{-1}}(0)
\sigma_{\beta{-1}}(1)\sigma_{\alpha\beta}(z_\infty)\rangle}=&
2\eta_{\beta}(-1){2\eta_\alpha-\eta_\beta}\qquad {\hbox{for}}\
\eta_\alpha< 1-\eta_\beta,\cr &
2(1-\eta_{\alpha})(-1){2\eta_\alpha-\eta_\beta} \qquad
{\hbox{for}}\ \eta_\alpha>1-\eta_\beta.}\eqn\france$$
The relevant Yukawa couplings are those for excited twist fields
that create normalized states. Consistently with the remarks
following \ma\ we should define the Yukawa coupling
$$YE_{\alpha{-1},\beta{-1},\alpha\beta}={1\over2}\eta{-1/2}_\beta
(1-\eta_\alpha){-1/2}(-1){\eta_\beta-2\eta_\alpha}
\langle{\tilde\tau}_{\alpha{-1}}(0){{\tilde\tau}'}_{\beta{-1}}(1)
\sigma_{\alpha\beta}(z_\infty)\rangle\eqn\be$$
so that
$$\eqalign{{YE_{\alpha{-1},\beta{-1},\alpha\beta}\over
\langle\sigma_{\alpha{-1}}(0){\sigma}_{\beta{-1}}(1)
\sigma_{\alpha\beta}(z_\infty)\rangle}&=
\sqrt{\eta_\beta\over (1-\eta_\alpha)},\qquad
\eta_\alpha< 1-\eta_\beta,\cr
&\sqrt{(1-\eta_\alpha)\over\eta_\beta},\qquad
\eta_\alpha>1-\eta_\beta.}\eqn\beirut$$

This result can be used to obtain the Yukawa couplings for
${\bf Z}_M\times {\bf Z}_N$ orbifolds when some of the twisted
sector states involved are excited states from the Yukawa
couplings between twisted sector ground states evaluated elsewhere
[\twelve]. The twist dependent suppression factors that arise may be of
significance in obtaining the detailed pattern of quark and lepton
masses.
\vskip 1cm

\centerline{ACKNOWLEDGEMENTS}
This work was supported in part by S.E.R.C.
\vfill\eject
\centerline{\bf{Table Captions}}
Table 1: Point group elements and lattices for ${\bf Z}_M\times {\bf Z}_N$
orbifolds. The point group elements $\theta$ and $\omega$ which are
of the form $(e{2\pi i\eta_1},e{2\pi i\eta_2},e{2\pi
i\eta_3})$, are specified in the table by $(\eta_1,
\eta_2,\eta_3).$

Table 2: Possible excited three-point functions ${\Big({\cal
Z}_i3\Big)}_{excited}$ for the ${\bf Z}_3\times {\bf Z}_6$
orbifolds. The $\thetap\omegaq$ twisted sector is denoted
by $T_{pq}$. Factors in ${\Big({\cal
Z}_i3\Big)}_{excited}$ involving only
unexcited twist fields are not displayed, nor are two-point
functions involving excited twist fields, which can be normalized
to 1. The notation is a slight modification of that of Eq. 9 with
the twists $\etai_\alpha$, $\etai_\beta$ and
$\etai_\gamma$  displayed as subscripts.
\vfill\eject
\centerline{\bf{TABLE 1}} \vskip 0.5cm
\begintable
Point Group |$\theta$ |$\omega$ |Lattice\cr${\bf Z}_2\times
{\bf Z}_2$ |$(1,0,-1)/2$ |$(0,1,-1)/2$ |$SO(4)3$\cr${\bf Z}_3\times
{\bf Z}_3$ |$(1,0,-1)/3$ |$(0,1,-1)/3$ |$SU(3)3$\cr${\bf Z}_2\times
{\bf Z}_4$ |$(1,0,-1)/2$ |$(0,1,-1)/4$ |$SO(4)\times SO(5)2$\cr${\bf
Z}_4\times
{\bf Z}_4$ |$(1,0,-1)/4$ |$(0,1,-1)/4$ |$SO(5)3$\cr${\bf Z}_2\times
{\bf Z}_6$ |$(1,0,-1)/2$ |$(0,1,-1)/6$ |$SO(4)\times G_22$\cr
${\bf Z}_2\times
{\bf Z}'_6$ |$(1,0,-1)/2$ |$(1,1,-2)/6$ |$G_23$\cr
${\bf Z}_3\times
{\bf Z}_6$ |$(1,0,-1)/3$ |$(0,1,-1)/6$ |$SU(3)\times G_22$ \cr${\bf
Z}_6\times {\bf Z}_6$ |$(1,0,-1)/6$ |$(0,1,-1)/6$ |$G_23$
\endtable
\vskip 0.5cm
\centerline{\bf{TABLE 2}}
\vskip 0.5cm
\begintable
Yukawa Coupling|Possible excited three-point
functions\cr
$T_{01}T_{14}T_{21}$|
$\langle\sigma2_{1/6}{\tilde\tau}'{}2_{2/3}
{\tilde\tau}2_{1/6}\rangle$,\nr
|
$\langle{\tilde\tau}2_{1/6}{\tilde\tau}'{}2_{2/3}\sigma2_{1/6}\rangle$
\cr $T_{02}T_{13}T_{21}$|
$\langle{\tilde\tau}'{}3_{2/3}{\tilde\tau}3_{1/6}
\sigma3_{1/6}\rangle$,\nr |
$\langle{\tilde\tau}'{}3_{2/3}\sigma3_{1/6}{\tilde\tau}3_{1/6}\rangle$\cr
$T_{04}T_{11}T_{21}$|$\langle{\tilde\tau}'{}2_{2/3}
{\tilde\tau}2_{1/6} \sigma2_{1/6}\rangle$,
\nr | $\langle{\tilde\tau}'{}2_{2/3}\sigma2_{1/6}
{\tilde\tau}2_{1/6}\rangle$\cr
$T_{05}T_{10}T_{21}$|
$\langle{\tilde\tau}3_{1/6}{\tilde\tau}'{}3_{2/3}
\sigma3_{1/6}\rangle$,
\nr | $\langle\sigma3_{1/6}{\tilde\tau}'{}3_{2/3}
{\tilde\tau}3_{1/6}\rangle$
\cr $T_{10}T_{13}T_{13}$|
$\langle{\tilde\tau}'{}3_{2/3}{\tilde\tau}3_{1/6}
\sigma3_{1/6}\rangle$,\nr
| $\langle{\tilde\tau}'{}3_{2/3}\sigma3_{1/6}
{\tilde\tau}3_{1/6}\rangle$
\cr  $T_{11}T_{11}T_{14}$|
$\langle{\tilde\tau}2_{1/6}\sigma2_{1/6}
{\tilde\tau}'{}2_{2/3}\rangle$,\nr
| $\langle\sigma2_{1/6}{\tilde\tau}2_{1/6}
{\tilde\tau}'{}2_{2/3}\rangle$
\endtable
\vfill\eject
\refout
\end